\documentclass[12pt]{iopart}

\usepackage{graphicx}
\begin{document}

\title[What is there in the black box of dark energy: variable cosmological parameters or ...]{What is there in the black box of dark energy: variable cosmological parameters  or multiple (interacting) components?}

\author{J Grande$^1$, J Sol\`{a}$^{1,}$$^2$ and H \v{S}tefan\v{c}i\'{c}$^{1,}$$^3$ \footnote{Speaker. Talk given at IRGAC 2006, Barcelona, July 11-15 2006.}}

\address{$^1$ High Energy Physics Group, Dep. ECM,  Universitat de
Barcelona,\\ \hspace{0.2cm}
Diagonal 647, 08028 Barcelona, Catalonia, Spain}

\address{$^2$ C.E.R. for Astrophysics, Particle Physics and
Cosmology \footnote{Associated with Instituto de Ciencias del
Espacio-CSIC.}}

\address{$^3$ Theoretical Physics Division, Rudjer Bo\v{s}kovi\'{c}
Institute, P.O.B. 180, HR-10002 Zagreb, Croatia}

\ead{jgrande@ecm.ub.es, sola@ifae.es, shrvoje@thphys.irb.hr}

%

\begin{abstract}

The coincidence problems and other dynamical features of dark energy are studied in cosmological models with variable cosmological parameters and in models with the composite dark energy. It is found that many of the problems usually considered to be cosmological coincidences can be explained or significantly alleviated in the aforementioned models.

\end{abstract}


\section{Introduction}

The accelerated expansion of the present universe, well established by a number of cosmological observations \cite{Supernovae1,Supernovae2,WMAP3Y, LSS}, challenges our understanding of the world around us at the most fundamental level. The dynamics causing this accelerated expansion is still unknown, although the list of proposed models aiming at the explanation of this puzzle is anything but short. Some interesting directions of research of this phenomenon comprise the theories of modified gravity \cite{ModGrav} and braneworld models \cite{Brane}. Another prominent approach to the problem of the acceleration of the universe is the hypothesis of the existence of {\em dark energy}, a cosmic component with the negative pressure \cite{Copeland06}. Apart from being a very widespread concept in both the theoretical modeling and the analysis and parametrization of the observational data, it is also useful as an effective description of other approaches towards the solution of the acceleration puzzle.

Despite the arrival of a host of observational data of ever increasing quality and quantity, the insight into the fundamental nature of the dark energy component is still quite limited. The effects of dark energy are observable through its gravitational interaction at the cosmological distances which presently prevents us from looking into the finer details of dark energy. In a way, dark energy resembles a black box with hidden inner components and an observable signal. In such a situation, any peculiarity of this signal is invaluable for inferring the properties of dark energy. Our present picture of the universe indeed does provide some specific signals, which are usually called coincidences in the cosmological context. Namely, presently the energy densities of dark energy and nonrelativistic matter are of the same order of magnitude despite the fact that their scalings with the redshift are very different. A specific type of coincidence is that the present value of the dark energy equation of state parameter $w$ is presently close to -1 and that it has possibly crossed the $w=-1$ line at a small redshift. In this paper we show how these specific signals from the output of the dark energy black box can be interpreted in models with variable cosmological parameters or in models with the composite dark energy.

\section{Cosmology with variable cosmological parameters}

The motivation for the investigation of cosmological models with variable parameters comes from a number of approaches, e.g., from the renormalization group equations (RGEs) for parameters in quantum field theory (QFT) in curved spacetime \cite{nova2,BGHS,nova,Bauer} or in quantum gravity (QG) \cite{Reuter}. We consider a general class of these models irrespective of the origin of the variability of the parameters \cite{SS12,SS14}. The variability of the parameters is introduced at the level of the Einstein equations
\begin{equation}
R_{\mu \nu }-\frac{1}{2}g_{\mu \nu }R=8\pi
G\,T_{\mu\nu}+\,g_{\mu\nu}\Lambda \equiv 8\pi
G\,(T_{\mu\nu}+g_{\mu\nu}\,\rho_{\Lambda})\,, \label{EE}
\end{equation}
i.e., the parameters $\rho_{\Lambda}$ and/or $G$ in these equations are considered to be variable. The variability of these parameters is in general constrained by the generalized Bianchi identity
\begin{equation}\label{GBI}
\bigtriangledown^{\mu}\,\left[G\,(T_{\mu\nu}+g_{\mu\nu}\,\rho_{\Lambda})\right]=0\, ,
\end{equation}
which in the FRW metric acquires the following form:
\begin{equation}\label{BianchiGeneral}
\frac{d}{dt}\,\left[G(\rho+\rho_{\Lambda})\right]+3\,G\,H_{\Lambda}\,(\rho+p)=0\,.
\end{equation}
This expression connects the variation of $\rho_{\Lambda}$ and $G$ with the (non)conservation of the matter component $\rho$. Consequently, the nonconservation of $\rho$ may be compensated by the variability of $\rho_{\Lambda}$ and/or $G$. The expansion of the universe in a model with the variable parameters is, for a spatially flat universe, described by the Friedmann equation
\begin{equation}\label{CCpicture}
H_{\Lambda}^2=\frac{8\pi G}{3}(\rho+\rho_{\Lambda})\,.
\end{equation}
The expressions (\ref{BianchiGeneral}) and (\ref{CCpicture}) do not determine the dynamics of the model completely, i.e. some additional dynamical information must be separately provided by the model of the variability of parameters (such as the RGE from QFT in curved spacetime). Usually it is given in terms of other relevant cosmological quantities
\begin{equation}\label{varibleCCG}
\rho_{\Lambda}(z)=\rho_{\Lambda}(\rho(z),H(z),...)\,,\ \ \ \ \
G(z)=G(\rho(z),H(z),...)\,.
\end{equation}
The Friedmann equation can then be written in the form
\begin{equation}\label{HLambda}
H_{\Lambda}^2(z)=H^2_0\,\left[\Omega_M^0\,f_M(z;r)(1+z)^{\alpha}
+\Omega_{\Lambda}^0\,f_{\Lambda}(z;r)\right]\,,
\end{equation}
where $f_M$ and $f_{\Lambda}$ are the functions of redshift, $r$ denotes the parameters of the model (\ref{varibleCCG}), and $f_M(0;r)=f_{\Lambda}(0;r)=1$ is satisfied. The description of the dynamics of the universe given so far we call the Cosmological Constant (CC) picture.

It is possible to construct the equivalent dynamical description, which we call the Dark Energy (DE) picture (the corresponding energy fraction parameters are denoted by $\tilde{\Omega}$) . The equivalence of these two pictures, i.e. their matching, is ensured by imposing $H_{\Lambda}=H_{D}$, where the indices $\Lambda$ and $D$ are added to differ the expressions for $H$ in the two pictures. The DE picture contains two noninteracting components, the conserved nonrelativistic matter component $\rho_{s}$ and
 the conserved DE component $\rho_D(z)=\rho_D(0)\,\zeta(z)$ with
$\zeta(z)\equiv\,\exp\left\{3\,\int_0^z\,dz' \frac{1+w_{eff}(z')}{1+z'}\right\}$.
%
%
The matching of the two pictures $H_{\Lambda}=H_{D}$ in combination with (\ref{BianchiGeneral}) leads to
\begin{equation}\label{droro}
{(1+z)}\,\,d(\rho_s+\rho_D)=\alpha\,\left(\rho_s+\rho_D-\xi_{\Lambda} \right)\,{dz}\,,
\end{equation}
where $\xi_{\Lambda}(z)=\frac{G(z)}{G_0}\,\rho_{\Lambda}(z)$.
From (\ref{droro}) it follows that the RG-like equation for $\rho_{D}$ is
\begin{equation}\label{drdz}
\frac{d\rho_D(z)}{dz}=\alpha\,\frac{\rho_D(z)-\xi_{\Lambda}(z)}{1+z}\equiv\beta(\rho_D(z))\,.
\end{equation}
The EOS parameter $w_{eff}$ for the effective DE component is given by
\begin{equation}\label{we2}
w_{eff}(z)=-1+\frac{\alpha}{3}\,\left(1-\frac{\xi_{\Lambda}(z)}{\rho_D(z)}\right)\equiv-1+\epsilon(z)\, ,
\end{equation}
which indicates that the possible crossing of the CC boundary happens at the redshift $z^{*}$ at which $\rho_D(z^{*})=\xi_{\Lambda}(z^{*})$. Using this property of $z^{*}$, (\ref{drdz}) can be put into the following form:
\begin{equation}\label{IF2}
\rho_D(z)=\xi_{\Lambda}(z)-\left(1+z\right)^{\alpha}\,
\int_{z^{*}}^z\frac{dz'}{(1+z')^{\alpha}}\frac{d\xi_{\Lambda}(z')}{dz'}\,,
\end{equation}
from which one readily obtains the expression for the slope
\begin{equation}\label{dIF2}
\frac{d\rho_D(z)}{dz}=-\alpha\,\left(1+z\right)^{\alpha-1}
\int_{z^{*}}^z\frac{dz'}{(1+z')^{\alpha}}\frac{d\xi_{\Lambda}(z')}{dz'}\,.
\end{equation}
This expression reveals a very interesting characteristic of $\rho_D(z)$. When $\xi_{\Lambda}(z)$ is a monotonous function of $z$ (which is the case in practically all models in the literature), the signs of slopes of $\rho_D(z)$ and $\xi_{\Lambda}(z)$ are opposite for $z>z^{*}$. The signs of slopes of $\rho_D(z)$ and $\xi_{\Lambda}(z)$ are the same only for $z<z^{*}$. To illustrate how counterintuitive this result is, we can consider the case with $G=$const where $\xi_{\Lambda}=\rho_{\Lambda}$. In this case, we find that, for $z>z^{*}$, if $\rho_{\Lambda}$ is a decreasing (increasing) function of $z$, $\rho_{D}$ is an increasing (decreasing) function of $z$, a general and somewhat surprising result.

\begin{figure}
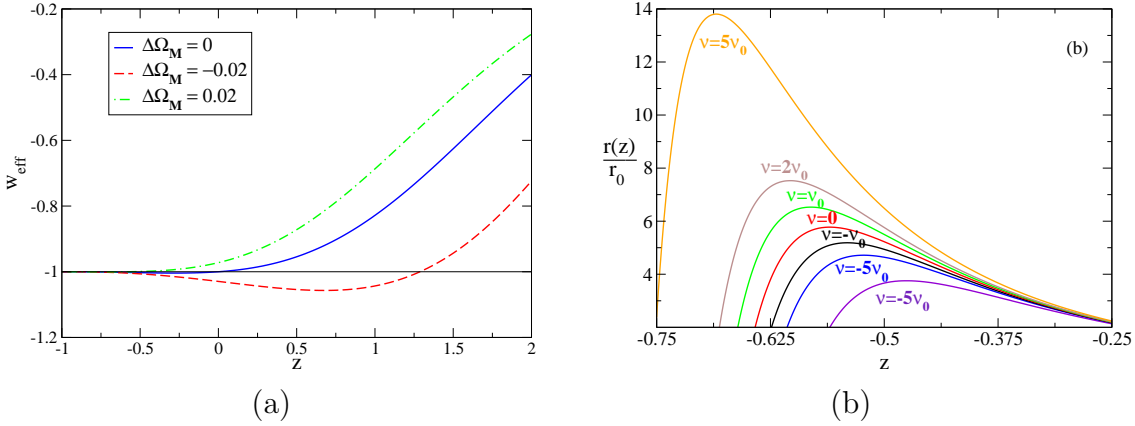



\begin{tabular}{cc}
     \resizebox{0.45\textwidth}{!}{\includegraphics{minusnu2z.eps}} &
     \hspace{0.3cm}
     \resizebox{0.45\textwidth}{!}{\includegraphics{ratcomb.eps}} \\
      (a) & (b)
    \end{tabular}

\caption{(a) The redshift dependence of $w_{eff}$ for the RG motivated model with $G=$const and $\rho_{\Lambda}=C_0+C_1 H^2$ \cite{RGTypeIa,RGTypeIa2,SS13}, for $\nu=-\nu_0=-0.026$, $\Omega_M=0.3$, $\Omega_{\Lambda}=0.7$ and various values of $\Delta \Omega_M=\Omega_M^0-\tilde{\Omega}_M^0$. (b) The plot of the ratio $r$ for $w_X=-1.85$, $\Omega_{\Lambda}=0.75$ and various values of $\nu$ \cite{GSS1}.}
\label{slika1}
\end{figure}

The final question to be answered is the following: what is the value of the characteristic redshift $z^{*}$? Using the generalized Bianchi identity (\ref{BianchiGeneral}) and the matching condition $H_{\Lambda}=H_D$, it is straightforward to obtain the following expression:
\begin{equation}\label{dzeta}
\frac{d\zeta(z)}{dz}=\frac{\alpha\,(1+z)^{\alpha-1}}{1-\tilde{\Omega}_M^0}
\left(\Omega_M^0\,f_M(z;r)-\tilde{\Omega}_M^0\right)\,.
\end{equation}
From the expression for $\zeta(z)$ it is easy to see that at $z^{*}$, where $w_{eff}(z)=-1$ vanishes, the derivative $\frac{d\rho_D(z)}{dz}$ becomes 0. Since the difference between $\Omega_M^0$ and $\tilde{\Omega}_M^0$ is small and $f_M(0;r)=1$ \cite{SS12}, the continuity of $f_M(z)$ guarantees that $z^{*}$ is close to 0. The value of $z^{*}$ may be positive, zero, or negative, which corresponds to the CC boundary crossing which happened in the past, which is happening now, or which will happen in the future, as nicely demonstrated in Fig. \ref{slika1}a. Whichever the value of $z^{*}$, it is close to 0, which guarantees that $w_{eff}$ is close to -1 at the present epoch. Therefore, the generic prediction of the models with variable cosmological parameters is that in their effective DE description the parameter $w_{eff}$ is presently close to -1, while it might have substantially differed from -1 in the past, i.e., these models may have a considerably different past behaviour from the $\Lambda$CDM.

\section{Composite dark energy}

The models with the composite dark energy, i.e. dark energy consisting of multiple, possibly interacting, components, provide an intriguing explanation of the coincidence problem \cite{GSS1,GSS2}. We consider a broad class of such models, which we call $\Lambda$XCDM models, in which DE consists of a (variable) CC component and an additional $X$ component, referred to as {\it cosmon}, i.e., $\rho_D=\rho_{\Lambda}+\rho_X$.
%
We assume that $\rho_D$ is conserved, although the DE components can interact mutually \cite{GSS1}. The total DE conservation equation
\begin{equation}\label{conslawDE2}
\dot{\rho}_{\Lambda}+\dot{\rho}_X+\,\alpha_X\,\rho_X\,H=0\,, \ \ \ \ \
\alpha_X \equiv 3(1+w_X)\,, \\\ p_{X}=w_X \rho_{X} \,,
\end{equation}
determines the dynamics of $\rho_X$ once the form of variability of $\rho_{\Lambda}$ is specified. For $\rho_{\Lambda}=C_0+C_1 H^2$, the solution can be obtained in the closed form for a general EOS of the cosmon component, $w_X(z)$  \cite{GSS1}. Here we present the results for $w_X=$const for which the system dynamics is determined by the following equations:
\begin{eqnarray}\label{seteq}
&&\frac{d\,\rho_X}{dz}+\frac{d\,\rho_{\Lambda}}{dz}=\frac{\alpha_X\,\rho_X}{1+z}\,,\nonumber\\
&&\frac{d\rho_{\Lambda}}{dz}=\frac{3\,\nu}{8\,\pi}\,M_P^2\,\frac{dH^2}{dz}\,,\nonumber\\
&&
\,\frac{dH^2}{dz}=\frac{8\,\pi\,G}{3}\,\frac{\alpha_m\,\rho_m+\alpha_X\,\rho_X}{1+z}+2\,H_0^2\,\Omega_K^0\,(1+z)\, ,
\end{eqnarray}
where $\nu$ is a parameter introduced in \cite{RGTypeIa}, being responsible for the running of the CC.
The solution of this system \cite{GSS1} shows that the $\Lambda$XCDM model can successfully describe many observed DE properties. One of the most interesting possibilities contained in the $\Lambda$XCDM model is the possibility that the expansion of the universe stops at some future $z$. The conditions for the occurrence of stopping are discussed in detail in \cite{GSS1}. The cases characterized by the stopping of the expansion also exhibit a very interesting behavior of $r(z)=\rho_D(z)/\rho_m(z)$. Namely, this ratio develops a maximum in the future and achieves the value of -1 at the stopping point. A typical redshift dependence of $r$ is depicted in Fig. \ref{slika1}b. Although the height of the $r(z)$ curve at its maximum can in principle acquire a very broad range of values, this example further illustrates a very interesting possibility that the maximum is {\it low}, i.e., that $r(z_{max})$ is comparable with the present value of the ratio, $r_0$. To further investigate this exciting possibility, we study how the following conditions constrain the parametric space of the model: (i) at nucleosynthesis, $r_N< 10\%$; (ii) the stopping occurs; (iii) $r(z_{max})/r_0 < 10$. This analysis reveals that the volume in the parametric space corresponding to this interesting scenario is {\it non-negligible} \cite{GSS1}. This finding indicates that the conditions (i)-(iii) can be satisfied without fine-tuning, i.e., that the scenario of the low $r(z_{max})$ is rather generic in the $\Lambda$XCDM.

The scenario with a low $r(z_{max})$ answers, or at least significantly ameliorates the situation with the coincidence problem. Namely, within this scenario, from the onset of acceleration till the stopping, $r(z)$ remains bounded (as opposed to the $\Lambda$CDM model) and of the same order as the present value of this ratio. This fact puts the coincidence problem into a completely different perspective because the present value $r_0$ is by no means special. Moreover, in this scenario, $r_0$ is a {\it characteristic} value of $r(z)$, i.e., in the entire evolution from the onset of acceleration until the stopping, the ratio $r(z)$ is of the same order of magnitude as $r_0$.

Another interesting class of $\Lambda$XCDM models, called the type II $\Lambda$XCDM models and recently studied in \cite{GSS2}, are the models with variable $G$ and $\rho_{\Lambda}$ and the conserved X component. Although in some aspects the dynamical features of these models, as well as the constraints on the model parameters, differ from the dynamics in the    $\Lambda$XCDM models presented above (the type I models), the type II models also allow the realization of the scenario with the stopping and the low maximum of $r(z)$ in a sizable volume of the parametric space. This fact shows that both the type I and the type II $\Lambda$XCDM models are capable of solving the coincidence problem. This further reinforces the claim that the possibility of the solution of the coincidence problem is inherent to  $\Lambda$XCDM models in general, i.e., it is not a model-dependent effect.

\section{Conclusions}

The special features of the universe with dark energy, usually referred to as coincidences, provide a useful testing ground for candidate mechanisms of the acceleration of the universe. We have shown that the models with variable cosmological parameters provide a natural explanation why the EOS of the associated effective DE should be presently close to -1, whereas it may still have a non-negligible redshift dependence. On the other hand, the $\Lambda$XCDM models with the composite dark energy provide the generic explanation of the $r=\rho_{D}/\rho_{m}$ coincidence problem. Namely, in parts of the parametric space of these models with non-negligible volume, $r$ remains of the order of $r_0$ during the entire future expansion of the universe. In conclusion, variable cosmological parameters or the composite DE are possible mechanisms which explain some very characteristic DE properties and therefore act as notable DE candidates.

{\it Acknowledgment.} This work was supported in part by the Ministerio de Educaci\'on
y Ciencia of Spain (MEC) and FEDER under project
2004-04582-C02-01, and also by the DURSI under 2005SGR00564. JG was
also supported by the MEC under BES-2005-7803. The work of HS was
financed by the MEC and he thanks the Dep. ECM of the UB for
hospitality. He is also supported by the Ministry of Science, Education and Sport of the Republic of Croatia.

\end{document}